\documentclass[namedreferences]{solarphysics}
\usepackage[optionalrh]{spr-sola-addons}
\usepackage{graphicx}
\usepackage{url}
\begin{document}
\begin{article}
\begin{opening}
\title{Spatially-resolved Energetic Electron Properties for the 21 May 2004 Flare from Radio Observations and 3D Simulations}
\author{A.A.~\surname{Kuznetsov}$^{1}$\sep
        E.P.~\surname{Kontar}$^{2}$}
\runningauthor{Kuznetsov and Kontar}
\runningtitle{Energetic electron properties}
\institute{$^{1}$ Institute of Solar-Terrestrial Physics, Irkutsk 664033, Russia,
                  e-mail: \url{a_kuzn@iszf.irk.ru}\\
           $^{2}$ School of Physics and Astronomy, University of Glasgow, Glasgow G12 8QQ, UK,
                  e-mail: \url{Eduard.Kontar@glasgow.ac.uk}}
\begin{abstract}
We investigate in detail the 21 May 2004 flare using simultaneous observations of the {\it Nobeyama Radioheliograph}, {\it Nobeyama Radiopolarimeters}, {\it Reuven Ramaty High Energy Solar Spectroscopic Imager} (RHESSI) and {\it Solar and Heliospheric Observatory} (SOHO). The flare images in different spectral ranges reveal the presence of a well-defined single flaring loop in this event. We have simulated the gyrosynchrotron microwave emission using the recently developed interactive IDL tool GX Simulator. By comparing the simulation results with the observations, we have deduced the spatial and spectral properties of the non-thermal electron distribution. The microwave emission has been found to be produced by the high-energy electrons ($>100$ keV) with a relatively hard spectrum ($\delta\simeq 2$); the electrons were strongly concentrated near the loop top. At the same time, the number of high-energy electrons near the footpoints was too low to be detected in the RHESSI images and spatially unresolved data. The SOHO {\it Extreme-ultraviolet Imaging Telescope} images and the low-frequency microwave spectra suggest the presence of an extended ``envelope'' of the loop with lower magnetic field. Most likely, the energetic electron distribution in the considered flare reflects the localized (near the loop top) particle acceleration (injection) process accompanied by trapping and scattering.
\end{abstract}
\end{opening}

\section{Introduction}
Energetic electrons play a key role in solar flares and therefore knowing their distributions is highly important for better understanding the flare mechanisms and verifying the flare models. A lot of information ({\it e.g.}, the electron spectra, energetics, spatial distribution) can be inferred from the hard X-ray observations. However, at energies above $\sim 50$ keV, we usually see only hard X-ray emission from the footpoints of the flaring loops, which are located far from the particle acceleration sites; the electron number and spectra in the footpoints thus could be strongly affected by the propagation and trapping processes. At relatively low energies (up to a few tens of keV), the electrons in the corona can also be studied using X-ray imaging (see, {\it e.g.}, \opencite{kon11}; \opencite{guo12}; \opencite{jef13}). In particular, the observations by \inlinecite{kon11}, \inlinecite{bia11}, and \inlinecite{bia12} have indicated that the particle propagation in the flaring loops could be strongly affected by magnetohydrodynamic (MHD) turbulence. At higher energies ($\gtrsim 100$ keV), the coronal X-ray emission is usually too weak, so it can be observed only occasionally --- in partially occulted events where the bright footpoints are not visible \cite{kru08b,kru08a,kru10}.

On the other hand, the high-energy electrons in the solar corona can be studied using radio observations, because they produce intense gyrosynchrotron emission in the microwave range. Diagnosing the energetic electrons (and other parameters of solar flares) from the microwave emission meets two main difficulties. Firstly, we need well-calibrated observations with high spatial, temporal and spectral resolutions. Secondly, the emission mechanism (even for the incoherent gyrosynchrotron radiation) is rather complicated and depends on many parameters. Therefore, recovering the emission source parameters from microwave
observations is a nontrivial task, which generally requires 3D simulations. Such simulations have been performed by, {\it e.g.}, \inlinecite{pre92}, \inlinecite{kuc93}, \inlinecite{wan95}, \inlinecite{bas98}, \inlinecite{nin00}, \inlinecite{lee00}, and \inlinecite{tza08}. However, until now the number of such studies has been limited because precise gyrosynchrotron simulations for realistic 3D configurations tend to be very time-consuming.

An important step in improving the simulation tools was the development of the ``fast gyrosynchrotron codes'' \cite{fle10} that allow to compute the gyrosynchrotron emission parameters with high speed and accuracy. These codes have been used in the interactive IDL tool \texttt{GS Simulator} for 3D simulations of gyrosynchrotron emission from model magnetic tubes with a dipole magnetic field \cite{kuz11}. The next iteration of this simulation tool,  \texttt{GX Simulator}, uses realistic magnetic field configurations based on the extrapolation of observed photospheric magnetograms \cite{nit11a,nit11b,nit12}, which enables us to perform a quantitative comparison between the observations and the simulation results. By varying the model parameters and analyzing the simulation results, we can choose the set of parameters that provides the best fit to the observations. Since the automatic forward-fitting algorithms have not been implemented yet, the described diagnosing method can be effectively applied only to the events with the simplest structure.

In this work, we analyze the observations and perform simulations for the flare on 21 May 2004, in which the observations with different instruments indicate the presence of a well-resolved single flaring loop. The main purpose of the work is to reconstruct the spatial distribution of the energetic electrons along the loop and to determine the electron energy spectra in the corona. The observations are summarized in Section \ref{Observations}. In Section \ref{Simulations}, we present the 3D simulations of the microwave emission. The implications of the obtained results are discussed in Section \ref{Discussion}.
The conclusions are drawn in Section \ref{Conclusion}.

\begin{figure}
\centerline{\includegraphics{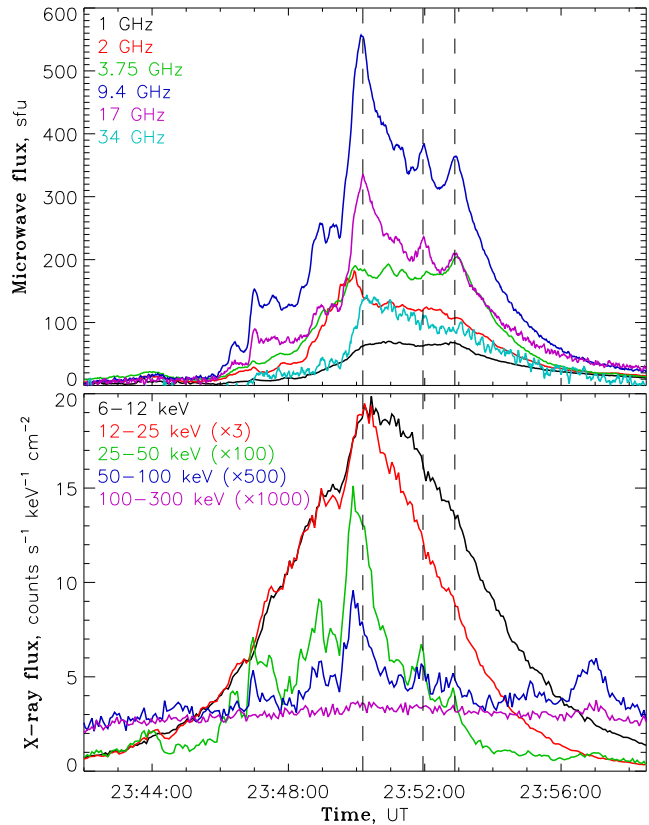}}
\caption{NoRP microwave (top) and RHESSI hard X-ray (bottom) lightcurves of 21 May 2004 flare.}
\label{FigLC}
\end{figure}

\section{Observations}\label{Observations}
We have selected a GOES M2.6 class flare that occurred on 21 May 2004 in the active region AR 10618 at S10E53. The microwave emission of this flare was observed by the {\it Nobeyama Radioheliograph} (NoRH) \cite{nak94} and {\it Radiopolarimeters} (NoRP); the hard X-ray emission was observed by the {\it Reuven Ramaty High Energy Solar Spectroscopic Imager} (RHESSI) \cite{lin02}. Figure \ref{FigLC} demonstrates the NoRP and RHESSI lightcurves (spatially integrated microwave and X-ray emissions). The microwave emission starts to increase after 23:46 UT and reaches its maximum at about 23:50 UT. After that, the intensity gradually decreases, although there are two additional weaker peaks at about 23:52 and 23:53 UT (see the 9.4 and 17 GHz lightcurves). The hard X-ray emission (especially in the 25--100 keV energy range) presents a similar behavior; the above mentioned three microwave peaks have well-visible counterparts in the 25--50 keV lightcurve. The main flare peak (at about 23:50 UT) is clearly visible in the 50--100 keV energy range as well. The X-ray emission at lower energies ($<25$ keV) varies more smoothly, while the higher-energy emission ($>100$ keV) presents no noticeable variation ({\it i.e.}, no flare-related signal) throughout the considered time interval.

\begin{figure}
\centerline{\includegraphics{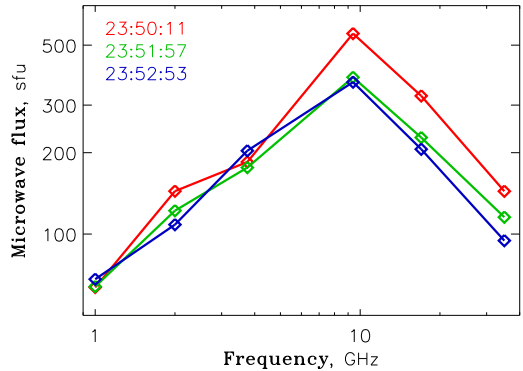}}
\caption{NoRP microwave spectra of the 21 May 2004 flare at different times (shown by vertical dashed lines in Figure \protect\ref{FigLC}).}
\label{FigObsSpectra}
\end{figure}

Figure \ref{FigObsSpectra} shows the NoRP microwave spectra (spatially unresolved) at three times corresponding to the strongest microwave emission peaks. The spectra have the shape typical of the nonthermal gyrosynchrotron emission mechanism; the emission peaks at 9.4 GHz. The shape of the spectra also suggests that the emission at the frequencies of $\gtrsim 17$ GHz is optically thin; the emission spectral index, $\delta_{\mathrm{MW}}$ (derived from the fluxes at 17 and 34 GHz), is about 1.15. According to the simulations of \inlinecite{dul82}, the spectral index of gyrosynchrotron emission is related to the spectral index of the emitting electrons $\delta$ as $\delta_{\mathrm{MW}}\simeq 0.90\delta-1.22$; this approximation yields a rather hard electron spectrum with $\delta\simeq 2.65$.

The spatially integrated RHESSI X-ray spectra of the considered flare (not shown in this paper) at energies below $\sim 100$ keV can be fitted by the thermal plus thick-target power-law model function with the temperature of the thermal component of about $2.3\times 10^7$ K. At higher energies ($>100$ keV), there was no reliable X-ray signal.

\begin{figure}
\centerline{\includegraphics{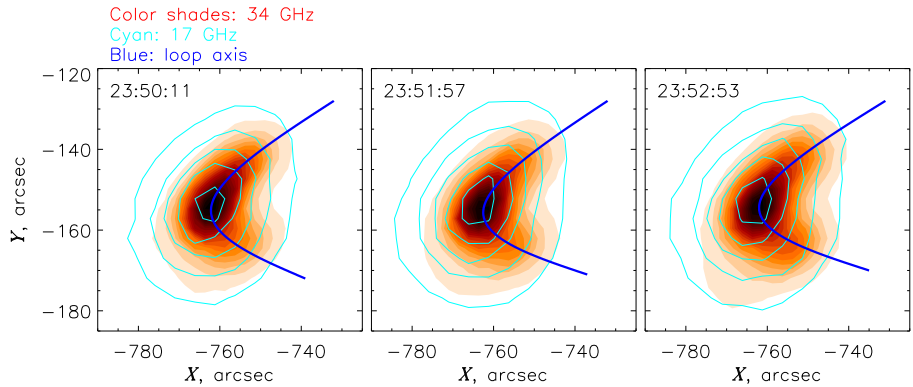}}
\caption{NoRH microwave images of 21 May 2004 flare at different times. The 17 GHz contours are drawn at 10, 30, 50, 70 and 90\% of the maximum intensity. In each panel, the suggested loop axis is shown by a blue line.}
\label{FigObsImages}
\end{figure}

Figure \ref{FigObsImages} shows the NoRH microwave images of the flare at three different times (corresponding to the microwave intensity peaks, like in Figure \ref{FigObsSpectra}). During the impulsive phase of the flare (23:47--23:59 UT), the microwave source structure remained nearly the same, despite of significant variations of the emission intensity. At 34 GHz, we see a well-defined loop-shaped structure with a distance between the footpoints of about $25''$ and a projected height of about $15''$. At 17 GHz, the emission source is larger, which is caused mainly by lower instrument resolution at this frequency (the NoRH beam widths at 17 and 34 GHz are about $18''$ and $9''$, respectively). As has been noted above, one of the aims of this work is to study the spatial distributions of the energetic electrons along the coronal loop; to do that, we have calculated firstly the 1D profiles of the microwave brightness temperature along the loop axis. The loop axis (at each time) was fitted by three-point cubic spline; the central point coincided with the 34 GHz intensity maximum and two other points were chosen to provide the best (``by-eye'') agreement of the resulting curve with the axis of the suggested magnetic loop, according to the 34 GHz image. The obtained axial lines are shown in Figure \ref{FigObsImages} and the corresponding 1D brightness temperature profiles are shown in Figure \ref{FigSimProfiles}. As one can notice from both the 2D images and 1D profiles, the emission is strongly peaked near the loop top.

\begin{figure}
\centerline{\includegraphics{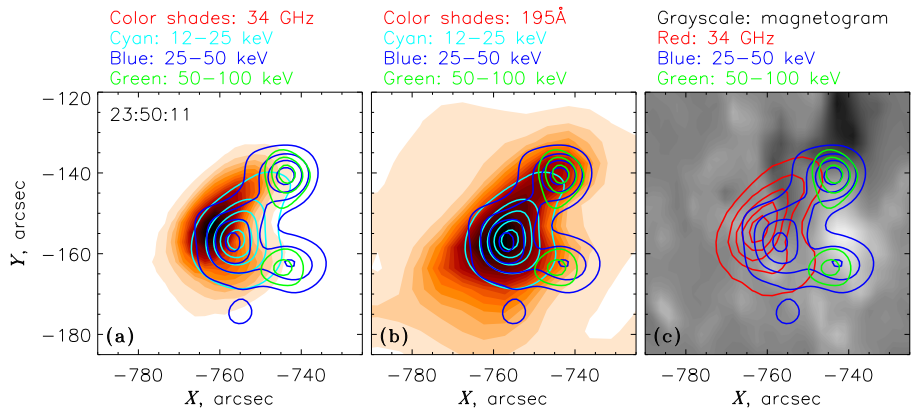}}
\caption{Images of 21 May 2004 flare at different wavelengths (at 23:50:11 UT). a) NoRH microwave and RHESSI hard X-ray images; b) SOHO/EIT EUV and RHESSI hard X-ray images; c) SOHO/MDI magnetogram, NoRH microwave and RHESSI hard X-ray images. The contours are drawn, relatively to the respective intensity maxima, at 10, 30, 50, 70 and 90\% for 34 GHz and 12-25 keV, at 30, 50, 70 and 90\% for 25-50 keV, and at 50, 70 and 90\% for 50-100 keV.}
\label{FigOverlays}
\end{figure}

The flare images in microwaves (34 GHz), hard X-rays ($12-100$ keV) and EUV (195~\AA) and the photospheric magnetogram are compared in Figure \ref{FigOverlays}; the time corresponding to the first and strongest microwave peak (23:50:11 UT) is chosen. The {\it Solar and Heliospheric Observatory Extreme-ultraviolet Imaging Telescope} (SOHO/EIT) EUV image and the {\it Solar and Heliospheric Observatory Michaelson Doppler Imager} (SOHO/MDI) magnetogram were recorded at 23:48:10 UT and 00:00:00 UT (of the next day), respectively, and were adjusted using the SolarSoft package to account for the SOHO location and differential rotation. At lower energies (12--25 keV), the X-ray image reveals the same loop-shaped structure as in microwaves; the emission is peaked at the loop top. At 25--50 keV, the visible loop becomes longer; in addition to the loop-top source, we see the loop footpoints\footnote{An additional weak compact source appearing at about ($-755''$, $-175''$) could be an artifact of the hard X-ray image reconstruction.}. At 50--100 keV, only the footpoints are visible. At energies above $100$~keV, the X-ray flux is too low to produce images. The 195~{\AA} image agrees well with both the microwave and (even better) the X-ray images and reveals the same single-loop structure. However, in addition to the main EUV loop, there is a region of decreasing emission extended to the South-East. The same feature can be noticed in the microwave images at later stages of the flare (see Figure \ref{FigObsImages}). According to the magnetogram, the considered active region contained two spots with strong magnetic fields of opposite polarities, whose locations agree well with the footpoints of the loop in microwaves, X-rays, and EUV. Therefore, we conclude that the 21 May 2004 flare had a relatively simple single-loop structure while producing intense and well-observed emission at different wavelengths (most importantly, in microwaves), which makes this flare a good candidate for further investigation.

\section{Simulations of the Loop Structure}\label{Simulations}
To investigate the parameters of energetic electrons in the considered flare, we have performed a number of 3D simulations of the gyrosynchrotron microwave emission. We used the recently developed interactive IDL tool \texttt{GX Simulator} \cite{nit11a,nit11b,nit12}. This tool allows us: a) to create a 3D magnetic field model based on the extrapolation of a photospheric magnetogram; b) to select a magnetic field line and coronal magnetic flux tube of interest; c) to populate that magnetic tube with thermal and nonthermal electrons with specified spatial, energy and pitch-angle distributions; d) to calculate 2D maps of the resulting gyrosynchrotron emission at different frequencies by integration of the radiation transfer equation along the lines-of-sight (using the fast gyrosynchrotron codes of \opencite{fle10}). The calculated maps (at 17 and 34 GHz) were then convolved with the corresponding NoRH response functions to obtain the images as they would be observed by the radioheliograph. The 1D brightness temperature profiles along the loop axis were calculated using those convolved images and the same axial lines as for the observed images.

\begin{figure}
\centerline{\begin{tabular}{cc}
\resizebox{5.26cm}{!}{\includegraphics{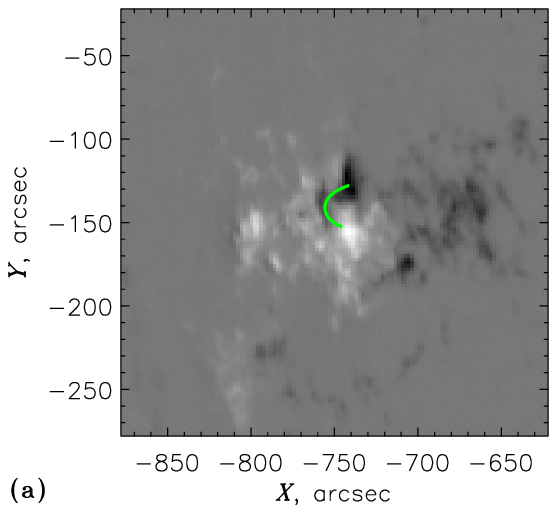}} &
\resizebox{5.26cm}{!}{\includegraphics{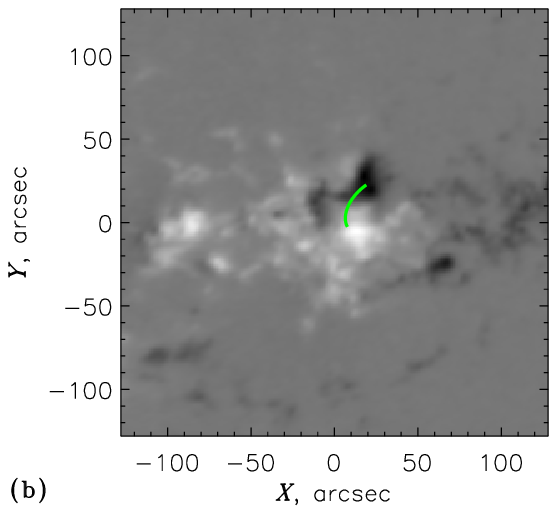}}\\
\makebox[10.5mm][l]{\small (c)}\resizebox{4.2cm}{!}{\includegraphics{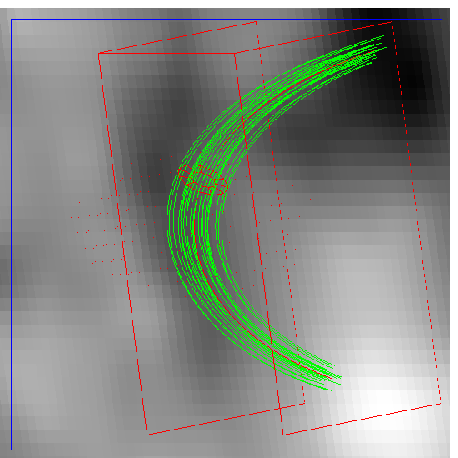}} &
\makebox[10.5mm][l]{\small (d)}\resizebox{4.2cm}{!}{\includegraphics{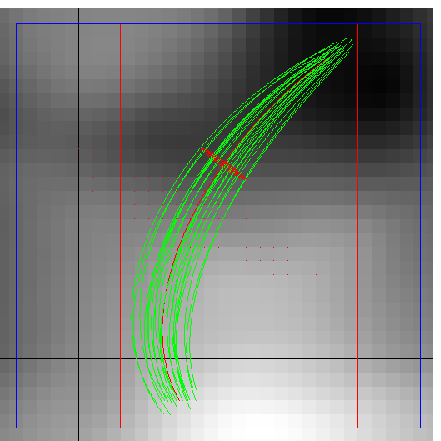}}
\end{tabular}}
\caption{(a) The observed photospheric magnetogram of AR 10618 at 23:50:11 UT on 21 May 2004. (b) The observed magnetogram rotated to the disk center ({\it i.e.}, as if it was observed from above). The axis of the selected ``active'' magnetic tube is shown by the green line. (c)-(d) Zoomed-in views of the panels (a) and (b), respectively, with the selected magnetic tube (screenshots from the \texttt{GX Simulator}). The image sizes in panels (c) and (d) are $37.5''\times 37.5''$ and $35.5''\times 35.5''$, respectively. The red box is the 3D box bounding the magnetic tube (in the heliographyc coordinates), while the red ellipses mark the loop top ({\it i.e.}, the cross-section corresponding to the lowest magnetic field).}
\label{FigMDI}
\end{figure}

The magnetic field was reconstructed using the linear force-free extrapolation, $\nabla\times\mathbf{B}=\alpha\mathbf{B}$ (see, {\it e.g.}, \opencite{nak72}; \opencite{see78}); the corresponding numerical code was provided by Costa (2013, private communication). This code requires knowing the vertical ($B_z$) component of the photospheric magnetic field, while SOHO/MDI measures only the line-of-sight component of the magnetic field and the considered active region was located rather far from the center of the solar disk. Therefore, some additional transformations have been applied (see Figure \ref{FigMDI}). Firstly, the selected area of the magnetogram (containing the active region, see Figure \ref{FigMDI}a) was rotated to the disk center under the assumption that the magnetic field at the photospheric level was perpendicular to the surface, {\it i.e.}, the resulting magnetic field (Figure \ref{FigMDI}b) was calculated as $B_z=B=B_{\mathrm{SOHO}}/\cos\vartheta$, where $\vartheta$ is the angular distance from the disk center to a given SOHO/MDI pixel. The resulting photospheric magnetic field varied from $-1900$ G (in the northern spot) to $+2200$ G (in the southern spot). Then we performed the extrapolation (Figure \ref{FigMDI}d); the parameter $\alpha$ was chosen to provide the best fit of the calculated magnetic field lines to the observed microwave loop (see below). Finally, the obtained 3D model of the magnetic field was rotated back to the original location of the active region (Figure \ref{FigMDI}c).

\begin{figure}
\centerline{\includegraphics{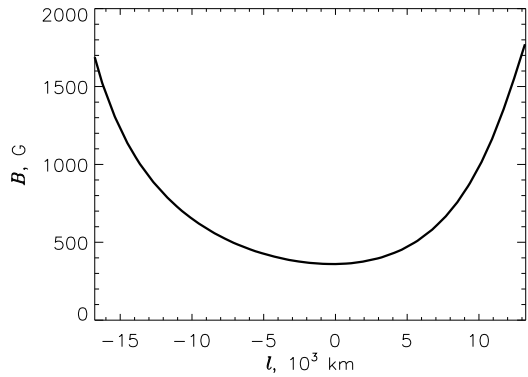}}
\caption{Magnetic field strength at the loop axis {\it vs.} the coordinate along the loop ($l=0$ corresponds to the minimum magnetic field).}
\label{FigBprofile}
\end{figure}

Using the extrapolated magnetic field, we have searched for the  magnetic tube whose axis (as seen from the Earth) coincided with the axial line of the observed microwave (34 GHz) loop. Since the NoRH positioning is not always sufficiently accurate, we have focused on reproducing the microwave loop shape and dimensions, while applying an appropriate (variable) shift to the observed images. Another variable parameter at this stage was the linear force-free parameter $\alpha$. The best agreement with the observations was obtained for $\alpha\simeq -1.4\times 10^{-10}$ $\textrm{cm}^{-1}$ and the chosen magnetic tube is shown
in Figures \ref{FigMDI}c-d. Figure \ref{FigBprofile} demonstrates the dependence of the magnetic field on the coordinate along the magnetic tube; the minimum magnetic field strength (at the loop top) is about $B_0\simeq 360$ G. The same magnetic tube (with tiny corrections due to the solar rotation) was used for all considered times.

The spatial distribution of the energetic electrons in the magnetic tube was modeled as $n_{\mathrm{b}}(l, r)=n_{\mathrm{b}0}(l)\varphi(r)$, where $l$ is the coordinate along the tube axis (with $l=0$ corresponding to the loop top, {\it i.e.}, to the point of the lowest magnetic field) and $r$ is the distance from the tube axis. The actual thickness of the flaring loop is hard to estimate, because of the insufficient instrument resolution (the observed thickness is very similar to the NoRH beam width). However, since the microwave emission at high frequencies is optically thin, the cross-sectional radius of the magnetic tube is not very important for the simulations, provided that this radius is sufficiently small --- in this case, the emission intensity at the tube axis should be simply proportional to the number of energetic electrons at a given cross-section. We modeled the electron distribution across the magnetic tube as $\varphi(r)=\exp(-r^2/a^2)$, with $a\simeq 730$ km ($\sim 1''$). Consequently, the energetic electron density at the tube axis $n_{\mathrm{b}0}(l)$ (see below) could be determined only with accuracy up to a constant factor depending on $a$.

\begin{table}
\caption{Parameters of the energetic electron distribution used in the model to fit the radio data. $n_{\max}$ is the maximum electron concentration, $l_0$ is the shift of the electron distribution relative to the loop top, the lengths $d_1$, $d_2$, $d_3$, $d_4$, $l_{12}$ and $l_{34}$ control the spatial distribution modeled by Equation (\protect\ref{spd}) and the resulting density of energetic electrons is presented in Figure \protect\ref{FigDensity}. $N_{\mathrm{total}}$ is the total number of the energetic electrons with $E\ge 60$ keV in the loop. $\Delta_{17}$ and $\Delta_{34}$ are the maximum relative differences between the observed and simulated brightness profiles at two frequencies.}
\label{TabParams}
\renewcommand{\tabcolsep}{28pt}
\begin{tabular}{lccc}
\hline
Time, UT & 23:50:11 & 23:51:57 & 23:52:53\\
\hline
$n_{\max}$, $\textrm{cm}^{-3}$ & $1.28\times 10^6$    & $4.45\times 10^5$    & $5.35\times 10^5$    \\
$\delta$                  & 2.10                 & 1.85                 & 1.95                 \\
$l_0$, km                 & -3220                & -3150                & -3290                \\
$l_{12}$, km              & -6000                & -5390                & -4500                \\
$l_{34}$, km              & 6600                 & 5990                 & 6300                 \\
$d_1$, km                 & 11\,500              & 29\,900              & 30\,000              \\
$d_2$, km                 & 4290                 & 3740                 & 3300                 \\
$d_3$, km                 & 6000                 & 4910                 & 4840                 \\
$d_4$, km                 & 12\,000              & 15\,000              & 20\,000              \\
$N_{\mathrm{total}}$      & $2.29\times 10^{31}$ & $7.27\times 10^{30}$ & $8.61\times 10^{30}$ \\
$\Delta_{17}$             & 0.107                & 0.057                & 0.096                \\
$\Delta_{34}$             & 0.082                & 0.089                & 0.029                \\
\hline
\end{tabular}
\end{table}

\begin{figure}
\centerline{\includegraphics{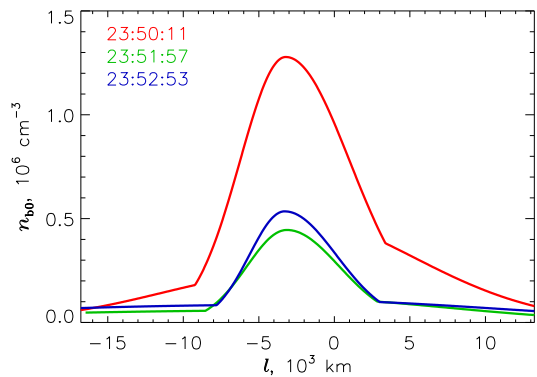}}
\caption{Spatial distributions of the energetic electrons $n_{\mathrm{b}0}(l)$ along the magnetic loop axis $l$ using the parameters from Table \protect\ref{TabParams}. Note that the peak of the distribution is shifted by $l_0$.}
\label{FigDensity}
\end{figure}

Our aim is to reproduce the observed 1D profiles of the microwave brightness temperature along the loop axis. To do this, we have tried different spatial distributions of the energetic electrons along the magnetic tube $n_{\mathrm{b}0}(l)$. As one may expect from radio images, the energetic electrons should be concentrated near the loop top. However, we have discovered that a simple ({\it e.g.}, exponential or gaussian) spatial distribution does not agree with the observations. Firstly, the electron distribution should have a sharp peak near the loop top, while near the footpoints it should become much smoother (that is, closer to a uniform one). Secondly, the electron distributions in different legs of the flaring loop seem to be asymmetric. Hence, we described the spatial distributions of energetic electrons along the tube axis by a model function
consisting of four Gaussians:

\begin{equation}\label{spd}
n_{\mathrm{b}0}(l)=n_{\max}\left\{\begin{array}{rl}
A_1\exp[-(l-l_0)^2/d_1^2], & l-l_0\le l_{12},\\[2pt]
\exp[-(l-l_0)^2/d_2^2], & l_{12}<l-l_0\le 0,\\[2pt]
\exp[-(l-l_0)^2/d_3^2], & 0<l-l_0\le l_{34},\\[2pt]
A_4\exp[-(l-l_0)^2/d_4^2], & l-l_0>l_{34},
\end{array}\right.
\end{equation}
where $n_{\max}$ is the maximum electron concentration, $l_0$ is the shift of the electron distribution relative to the loop top, the lengths $d_1$, $d_2$, $d_3$, $d_4$, $l_{12}$ and $l_{34}$ control the shape of the distribution, and the factors $A_1$ and $A_4$ are chosen to make the function $n_{\mathrm{b}0}(l)$ continuous. The parameters are given in Table \ref{TabParams}. Note that the problem of inversion of the solar radio images, obviously, has no unique solution. Therefore the model function given by Equation (\ref{spd}) is just one of the possibilities; it was chosen due to its (relative) simplicity and the easiness of implementation. The total number of energetic electrons in the active region is given by
\begin{equation}\label{Ntotal}
N_{\mathrm{total}}=2\pi\int\limits_0^{\infty}\varphi(r)r\,\textrm{d}r\int\limits_{l_{\min}}^{l_{\max}}n_{\mathrm{b}0}(l)\,\textrm{d}l,
\end{equation}
where $l_{\min}$ and $l_{\max}$ correspond to the points where the magnetic tube crosses the photosphere.

The energetic electrons are assumed to have a power-law energy distribution: $f(E)\sim E^{-\delta}$ for $E\ge 60$ keV (since the gyrosynchrotron radiation is produced mainly by high-energy electrons and the low-energy cutoff is not very important). The pitch-angle electron distribution was assumed to be isotropic. Since we are interested mainly in high-frequency emission where the influence of the thermal plasma ({\it e.g.}, the Razin suppression) is negligible, the thermal plasma density was not included into the fitting procedure, {\it i.e.}, this parameter was assumed to be the same and relatively small ($n_0=5\times 10^9$ $\textrm{cm}^{-3}$) in all the simulations.

We have performed a number of simulations, varying the parameters of the model function given by Equation (\ref{spd}) as well as the electron power-law index $\delta$. The aim was to obtain the best agreement between the simulated and observed 1D distributions of the microwave brightness temperature along the loop axis at both NoRH frequencies. Namely, the agreement (at a given emission frequency) was characterized by the value
\begin{equation}
\Delta=\frac{\max|T_{\mathrm{obs}}(s)-T_{\mathrm{mod}}(s)|}{\max T_{\mathrm{obs}}(s)},
\end{equation}
where $T_{\mathrm{obs}}(s)$ and $T_{\mathrm{mod}}(s)$ are respectively the observed and simulated brightness temperatures at the loop axis, and all the possible values of the coordinate $s$ along the loop axis are considered. The fitting was to minimize the value $\Delta_{17}+\Delta_{34}$, which implies minimizing the differences between the simulations and the observations at both 17 GHz ($\Delta_{17}$) and 34 GHz ($\Delta_{34}$). As mentioned above, at this stage of development of the simulation tool, adjustment of the model parameters and evaluation of the fit goodness were performed manually. The best fit parameters for three different times (corresponding to the microwave intensity peaks) are given in Table \ref{TabParams}. The corresponding spatial distributions of the energetic electrons along the magnetic tube axis are shown in Figure \ref{FigDensity}. The achieved values of $\Delta_{17}$ and $\Delta_{34}$ are also given in Table \ref{TabParams}; one can see that the observations and the simulations agree with an accuracy of about 10\% or better. Simulations with different model parameters suggest that possible variations of the electron distribution parameters near the loop top ($n_{\max}$, $\delta$, $l_0$, $l_{12}$, $l_{34}$, $d_2$, $d_3$) do not exceed $\sim 10$\%; the estimations of the distribution parameters near the footpoints are less reliable: the scales $d_1$ and $d_4$ may vary by a factor of up to $\sim 1.5$.

\begin{figure}
\centerline{\includegraphics{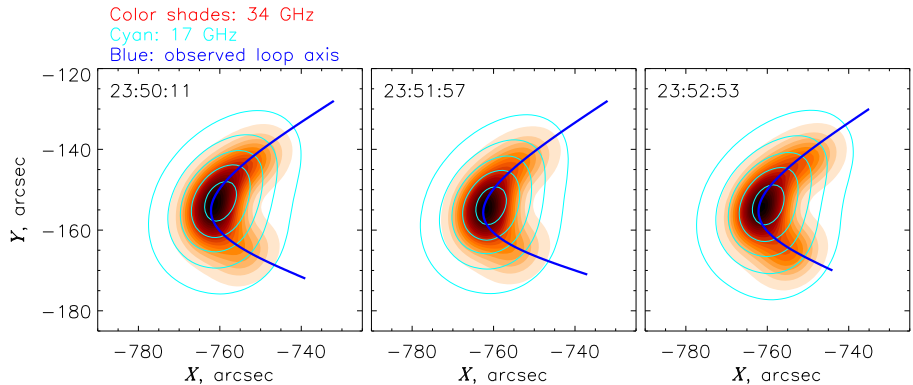}}
\caption{Simulated microwave images at different times. The axial lines are taken from the observations, {\it i.e.}, they are the same as in Figure \protect\ref{FigObsImages}.}
\label{FigSimImages}
\end{figure}

\begin{figure}
\centerline{\includegraphics{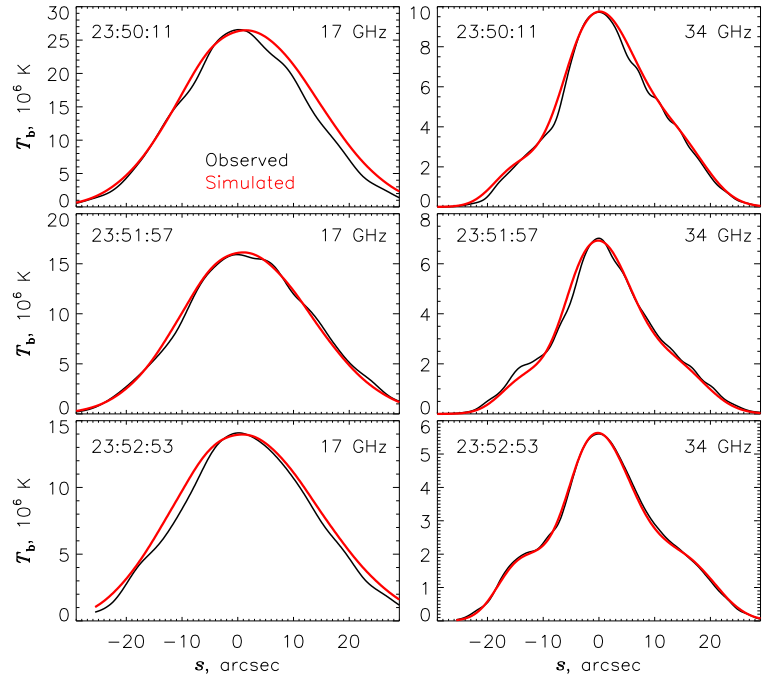}}
\caption{Observed and simulated 1D distributions of the microwave brightness temperature along the suggested loop axis at different times and frequencies.}
\label{FigSimProfiles}
\end{figure}

\begin{figure}
\centerline{\includegraphics{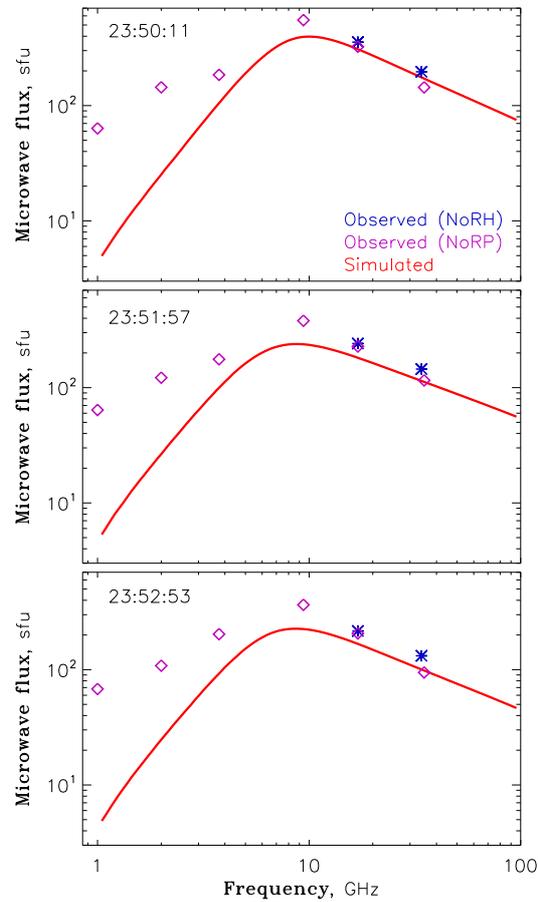}}
\caption{Observed and simulated total (spatially unresolved) emission spectra at different times.}
\label{FigSimSpectra}
\end{figure}

The computed 2D maps of the gyrosynchrotron emission (convolved with the NoRH response functions) are shown in Figure \ref{FigSimImages}, while the 1D distributions of the brightness temperature along the loop axis (the observed and simulated ones) are shown in Figure \ref{FigSimProfiles}. It is important that the simulations reproduce well the observed spectral index of the emission, {\it i.e.}, the ratio of intensities at 17 and 34 GHz. This result confirms, in particular, that an influence of the thermal plasma at the frequencies of $\gtrsim 17$ GHz can be neglected. The total (spatially integrated) emission spectra are presented in Figure \ref{FigSimSpectra}; the simulated spectra are shown by solid continuous lines, while the observations were available only at several frequencies. Note that there is a difference between the NoRH and NoRP data, caused by the calibration issues.

\section{Discussion}\label{Discussion}
As it can be noticed from Figure \ref{FigDensity}, the spatial distributions of energetic electrons at all considered times present similar features, despite significant variation in the particle concentration. It seems that the energetic electron population in the considered flare consisted of two components: one strongly peaked near the loop top component and a more homogeneous background. The strongly peaked component might be formed due to the localized (near the loop top) isotropic particle injection combined with the subsequent particle trapping in the inhomogeneous magnetic field, as shown, {\it e.g.}, in the numerical simulations by \inlinecite{mel06} and \inlinecite{gor07}. The weakly peaked (sometimes almost homogeneous) background might be formed as a result of scattering of the injected and trapped energetic particles, {\it e.g.}, due to small-scale magnetic turbulence \cite{kon11}. It is highly likely that the two mentioned electron components have different energy spectra, but currently we cannot check this hypothesis because of insufficient instrument resolution. Temporal variations of the electron parameters (see Table \ref{TabParams}), most likely, reflect the processes of impulsive particle acceleration/injection and subsequent particle propagation along the flaring loop.

The fact that the energetic electrons were concentrated at the loop top can explain the absence (non-detection) of the high-energy ($>100$ keV) X-ray emission; {\it i.e.,} in the coronal part of the flaring loop, where the high-energy electrons were abundant, the plasma density is relatively low and hence the thin-target bremsstrahlung emission (which is proportional to plasma density) is weak. On the other hand, in the loop footpoints (where the plasma density is high), the number of the high-energy electrons is negligible.

It can be noticed from Table \ref{TabParams} that the peak of the spatial distribution of the energetic electrons is slightly shifted from the loop top ($l_0\simeq -3200$ km). This may indicate the location of the particle injection point. However, this shift might also be caused by instrumental effects, {\it e.g.}, by incorrect alignment of the magnetogram and the microwave images and/or by inaccuracy of the magnetic field extrapolation.

The inferred electron energy spectra are rather hard (see Table \ref{TabParams}). Note that the best-fit spectral indices differ from the estimations based on the approximation proposed by \inlinecite{dul82}, which is caused mainly by the magnetic field inhomogeneity in the emission source in our simulations (see Figure \ref{FigBprofile}); similar results have been  obtained by \inlinecite{kuz11} for the gyrosynchrotron emission from model magnetic loops with inhomogeneous magnetic field. In addition, the spectral indices given in Table \ref{TabParams} represent some approximate (average) values for the entire flaring loop, while, as noted above, the actual electron spectra could be spatially dependent.

The simulated spectra of the total (spatially integrated) emission (see Figure \ref{FigSimSpectra}) at the frequencies of 17 and 34 GHz demonstrate a good agreement with the NoRP and NoRH observations. This is not surprising, because the simulation parameters were chosen to fit the spatially resolved observations at those frequencies. At lower frequencies, the agreement is not so good: the observed spectrum is flatter than the computed one. Nevertheless, the simulations are able to reproduce approximately the location and magnitude of the spectral peak, which confirms the reliability of the obtained model. At even lower frequencies ($\lesssim 4$ GHz), the simulations cannot reproduce the observations: the observed emission intensities are much higher than the simulated ones. Although the emission source in our simulations is strongly inhomogeneous (see Figures \ref{FigBprofile} and \ref{FigDensity}), this inhomogeneity ({\it i.e.}, when we consider only the main flaring loop) cannot account for the observed spectral flattening. Note that an increase of the thermal plasma density in the flaring loop would make the simulated spectra at low frequencies even steeper \cite{kuz10}.

The difference between the observed and simulated spectra at the frequencies $\lesssim 4$ GHz might be caused by a contribution of plasma emission, which is typical of this frequency range. Alternatively, the additional low-frequency emission could be produced in another gyrosynchrotron source, {\it e.g.}, in another (smaller) flaring loop or in an extended ``envelope'' of the main loop (filled with energetic electrons, but with a relatively low magnetic field). A similar assumption has been proposed in the past to interpret the radio observations of several events (see, {\it e.g.}, \opencite{kle86}; \opencite{bru94}). As has been said in Section \ref{Observations}, the existense of such extended emission source can be suggested also from the EUV observations (see Figure \ref{FigOverlays}b) and, partially, from the 34 GHz NoRH observations (Figure \ref{FigObsImages}).

The presence of an extended envelope around the flaring loop seems to contradict the above conclusion that the 21 May 2004 flare had a simple single-loop structure. This fact highlights that, even if we see a simple source structure at some wavelengths, the actual structure of the active region may be much more complicated. Nevertheless, the loop envelope seems to affect only the low-frequency radio emission; thus the diagnostics based on the NoRH observations (and applied to the main flaring loop only) remains reliable.

As noted above, the presented simulations were performed (for simplicity) for an isotropic pitch-angle distribution of the energetic electrons. Simulations for the loss-cone distribution provided almost the same results, because in the considered event most of the energetic electrons were concentrated near the loop top, where the loss-cone angle is small and therefore the incoherent gyrosynchrotron radiation from the loss-cone distribution becomes similar to that from the isotropic electrons \cite{kuz11}.

\section{Conclusion}\label{Conclusion}
We have shown that spatially-resolved microwave observations together with 3D simulations can be an effective tool for diagnosing the energetic electrons in solar flares. By using the IDL program \texttt{GX Simulator}, varying the model parameters and comparing the simulation results with observations, it is possible to reconstruct the spatial distributions of energetic electrons in flaring loops and to estimate their energy spectrum and total number. For the 21 May 2004 flare, we have achieved a good agreement between the simulated microwave data and both the spatially resolved and unresolved observations.

On the other hand, the described diagnosing method still requires some additional data, besides the microwave observations --- namely, a 3D model of the magnetic field in the corona. Currently, this field is obtained using extrapolation of a photospheric magnetogram and therefore the simulation/diagnosing results are dependent on the extrapolation method used. However, we anticipate that this problem will be solved soon by using simultaneous multiwavelength imaging observations in the radio/microwave range with new or upgraded instruments (such as the {\it Chinese Solar Radioheliograph}, {\it Upgraded Siberian Solar Radio Telescope} and {\it Expanded Owens Valley Solar Array}); we expect that the new observations will enable us not only to compare and verify different magnetic field extrapolation techniques, but to perform independent measurements of the coronal magnetic field.

We have found that in the analyzed flare (21 May 2004), the energetic electrons were concentrated near the loop top. It seems that the energetic electron population consisted of two components: a strongly peaked (near the loop top) component and a more homogeneous ``background''; this spatial distribution might be formed due to a combination of the processes of particle acceleration, trapping and scattering. The X-ray emission at high energies ($>100$ keV) was below the detection level, despite of a relatively large total number of high-energy electrons; this contradiction can be explained by the fact that most of the energetic electrons are trapped in the coronal part of the flaring loop, where they do not produce a significant X-ray emission. The microwave and EUV observations also indicate that, besides the main flaring loop, the active region might contain a more extended gyrosynchrotron emission source filled with energetic electrons but with a relatively low magnetic field.

\begin{acks}
This work was supported in part by the Russian Foundation of Basic Research
(grants 12-02-00173, 12-02-91161, 13-02-10009 and 13-02-90472) and by a Marie Curie International Research Staff Exchange Scheme "Radiosun" (PEOPLE-2011-IRSES-295272) and STFC consolidated grant. The authors thank Natasha Jeffrey for help to improve the manuscript.
\end{acks}

\end{article}

\begin{thebibliography}{30}
\ifx \bisbn   \undefined \def \bisbn  #1{ISBN #1}\fi
\ifx \binits  \undefined \def \binits#1{#1}\fi
\ifx \bauthor  \undefined \def \bauthor#1{#1}\fi
\ifx \batitle  \undefined \def \batitle#1{#1}\fi
\ifx \bjtitle  \undefined \def \bjtitle#1{\textit{#1}}\fi
\ifx \bvolume  \undefined \def \bvolume#1{\textbf{#1}}\fi
\ifx \byear  \undefined \def \byear#1{#1}\fi
\ifx \bissue  \undefined \def \bissue#1{#1}\fi
\ifx \bfpage  \undefined \def \bfpage#1{#1}\fi
\ifx \blpage  \undefined \def \blpage #1{#1}\fi
\ifx \burl  \undefined \def \burl#1{\textsf{#1}}\fi
\ifx \href  \undefined \def \href#1#2{\textsf{#2}}\fi
\ifx \doiurl  \undefined \def
  \doiurl#1{\href{http://dx.doi.org/#1}{\textsf{#1}}}\fi
\ifx \betal  \undefined \def \betal{\textit{et al.}}\fi
\ifx \binstitute  \undefined \def \binstitute#1{#1}\fi
\ifx \bctitle  \undefined \def \bctitle#1{#1}\fi
\ifx \beditor  \undefined \def \beditor#1{#1}\fi
\ifx \bpublisher  \undefined \def \bpublisher#1{#1}\fi
\ifx \bbtitle  \undefined \def \bbtitle#1{\textit{#1}}\fi
\ifx \bedition  \undefined \def \bedition#1{#1}\fi
\ifx \bseriesno  \undefined \def \bseriesno#1{\textbf{#1}}\fi
\ifx \blocation  \undefined \def \blocation#1{#1}\fi
\ifx \bsertitle  \undefined \def \bsertitle#1{\textit{#1}}\fi
\ifx \bsnm \undefined \def \bsnm#1{#1}\fi
\ifx \bsuffix \undefined \def \bsuffix#1{#1}\fi
\ifx \bparticle \undefined \def \bparticle#1{#1}\fi
\ifx \barticle \undefined \def \barticle#1{}\fi
\ifx \botherref \undefined \def \botherref#1{}\fi
\ifx \url \undefined \def \url#1{\textsf{#1}}\fi
\ifx \bchapter \undefined \def \bchapter#1{}\fi
\ifx \bbook \undefined \def \bbook#1{}\fi
\ifx \bcomment \undefined \def \bcomment#1{#1}\fi
\ifx \oauthor \undefined \def \oauthor#1{#1}\fi
\ifx \citeauthoryear \undefined \def \citeauthoryear#1{#1}\fi
\def \endbibitem {}
\ifx \bconflocation  \undefined \def \bconflocation#1{#1} \fi

\bibitem[\protect\citeauthoryear{{Bastian}, {Benz}, and {Gary}}{1998}]{bas98}
\begin{barticle}
\bauthor{\bsnm{{Bastian}}, \binits{T.S.}},
\bauthor{\bsnm{{Benz}}, \binits{A.O.}},
\bauthor{\bsnm{{Gary}}, \binits{D.E.}}:
\byear{1998},
\batitle{{Radio Emission from Solar Flares}}.
\bjtitle{Ann. Rev. Astron. Astrophys.}
\bvolume{36},
\bfpage{131}\,--\,\blpage{188}.
doi:\doiurl{10.1146/annurev.astro.36.1.131}.
\end{barticle}
\endbibitem

\bibitem[\protect\citeauthoryear{{Bian}, {Emslie}, and {Kontar}}{2012}]{bia12}
\begin{barticle}
\bauthor{\bsnm{{Bian}}, \binits{N.}},
\bauthor{\bsnm{{Emslie}}, \binits{A.G.}},
\bauthor{\bsnm{{Kontar}}, \binits{E.P.}}:
\byear{2012},
\batitle{{A Classification Scheme for Turbulent Acceleration Processes in Solar
  Flares}}.
\bjtitle{Astrophys. J.}
\bvolume{754},
\bfpage{103}.
doi:\doiurl{10.1088/0004-637X/754/2/103}.
\end{barticle}
\endbibitem

\bibitem[\protect\citeauthoryear{{Bian}, {Kontar}, and
  {MacKinnon}}{2011}]{bia11}
\begin{barticle}
\bauthor{\bsnm{{Bian}}, \binits{N.H.}},
\bauthor{\bsnm{{Kontar}}, \binits{E.P.}},
\bauthor{\bsnm{{MacKinnon}}, \binits{A.L.}}:
\byear{2011},
\batitle{{Turbulent cross-field transport of non-thermal electrons in coronal
  loops: theory and observations}}.
\bjtitle{Astron. Astrophys.}
\bvolume{535},
\bfpage{A18}.
doi:\doiurl{10.1051/0004-6361/201117574}.
\end{barticle}
\endbibitem

\bibitem[\protect\citeauthoryear{{Bruggmann} \textit{et~al.}}{1994}]{bru94}
\begin{barticle}
\bauthor{\bsnm{{Bruggmann}}, \binits{G.}},
\bauthor{\bsnm{{Vilmer}}, \binits{N.}},
\bauthor{\bsnm{{Klein}}, \binits{K.-L.}},
\bauthor{\bsnm{{Kane}}, \binits{S.R.}}:
\byear{1994},
\batitle{{Electron trapping in evolving coronal structures during a large
  gradual hard X-ray/radio burst}}.
\bjtitle{Solar Phys.}
\bvolume{149},
\bfpage{171}\,--\,\blpage{193}.
doi:\doiurl{10.1007/BF00645188}.
\end{barticle}
\endbibitem

\bibitem[\protect\citeauthoryear{{Dulk} and {Marsh}}{1982}]{dul82}
\begin{barticle}
\bauthor{\bsnm{{Dulk}}, \binits{G.A.}},
\bauthor{\bsnm{{Marsh}}, \binits{K.A.}}:
\byear{1982},
\batitle{{Simplified expressions for the gyrosynchrotron radiation from mildly
  relativistic, nonthermal and thermal electrons}}.
\bjtitle{Astrophys. J.}
\bvolume{259},
\bfpage{350}\,--\,\blpage{358}.
doi:\doiurl{10.1086/160171}.
\end{barticle}
\endbibitem

\bibitem[\protect\citeauthoryear{{Fleishman} and {Kuznetsov}}{2010}]{fle10}
\begin{barticle}
\bauthor{\bsnm{{Fleishman}}, \binits{G.D.}},
\bauthor{\bsnm{{Kuznetsov}}, \binits{A.A.}}:
\byear{2010},
\batitle{{Fast Gyrosynchrotron Codes}}.
\bjtitle{Astrophys. J.}
\bvolume{721},
\bfpage{1127}\,--\,\blpage{1141}.
doi:\doiurl{10.1088/0004-637X/721/2/1127}.
\end{barticle}
\endbibitem

\bibitem[\protect\citeauthoryear{{Gorbikov} and {Melnikov}}{2007}]{gor07}
\begin{barticle}
\bauthor{\bsnm{{Gorbikov}}, \binits{S.P.}},
\bauthor{\bsnm{{Melnikov}}, \binits{V.F.}}:
\byear{2007},
\batitle{{The numerical solution of the Fokker-Planck equation for modeling of
  particle distribution in solar magnetic traps}}.
\bjtitle{Matem. Mod.}
\bvolume{19}(\bissue{2}),
\bfpage{112}\,--\,\blpage{122}.
\end{barticle}
\endbibitem

\bibitem[\protect\citeauthoryear{{Guo} \textit{et~al.}}{2012}]{guo12}
\begin{barticle}
\bauthor{\bsnm{{Guo}}, \binits{J.}},
\bauthor{\bsnm{{Emslie}}, \binits{A.G.}},
\bauthor{\bsnm{{Kontar}}, \binits{E.P.}},
\bauthor{\bsnm{{Benvenuto}}, \binits{F.}},
\bauthor{\bsnm{{Massone}}, \binits{A.M.}},
\bauthor{\bsnm{{Piana}}, \binits{M.}}:
\byear{2012},
\batitle{{Determination of the acceleration region size in a loop-structured
  solar flare}}.
\bjtitle{Astron. Astrophys.}
\bvolume{543},
\bfpage{A53}.
doi:\doiurl{10.1051/0004-6361/201219341}.
\end{barticle}
\endbibitem

\bibitem[\protect\citeauthoryear{{Jeffrey} and {Kontar}}{2013}]{jef13}
\begin{barticle}
\bauthor{\bsnm{{Jeffrey}}, \binits{N.L.S.}},
\bauthor{\bsnm{{Kontar}}, \binits{E.P.}}:
\byear{2013},
\batitle{{Temporal Variations of X-Ray Solar Flare Loops: Length, Corpulence,
  Position, Temperature, Plasma Pressure, and Spectra}}.
\bjtitle{Astrophys. J.}
\bvolume{766},
\bfpage{75}.
doi:\doiurl{10.1088/0004-637X/766/2/75}.
\end{barticle}
\endbibitem

\bibitem[\protect\citeauthoryear{{Klein}, {Trottet}, and {Magun}}{1986}]{kle86}
\begin{barticle}
\bauthor{\bsnm{{Klein}}, \binits{K.-L.}},
\bauthor{\bsnm{{Trottet}}, \binits{G.}},
\bauthor{\bsnm{{Magun}}, \binits{A.}}:
\byear{1986},
\batitle{{Microwave diagnostics of energetic electrons in flares}}.
\bjtitle{Solar Phys.}
\bvolume{104},
\bfpage{243}\,--\,\blpage{252}.
doi:\doiurl{10.1007/BF00159969}.
\end{barticle}
\endbibitem

\bibitem[\protect\citeauthoryear{{Kontar}, {Hannah}, and {Bian}}{2011}]{kon11}
\begin{barticle}
\bauthor{\bsnm{{Kontar}}, \binits{E.P.}},
\bauthor{\bsnm{{Hannah}}, \binits{I.G.}},
\bauthor{\bsnm{{Bian}}, \binits{N.H.}}:
\byear{2011},
\batitle{{Acceleration, Magnetic Fluctuations, and Cross-field Transport of
  Energetic Electrons in a Solar Flare Loop}}.
\bjtitle{Astrophys. J.}
\bvolume{730},
\bfpage{L22}.
doi:\doiurl{10.1088/2041-8205/730/2/L22}.
\end{barticle}
\endbibitem

\bibitem[\protect\citeauthoryear{{Krucker} \textit{et~al.}}{2008a}]{kru08b}
\begin{barticle}
\bauthor{\bsnm{{Krucker}}, \binits{S.}},
\bauthor{\bsnm{{Hurford}}, \binits{G.J.}},
\bauthor{\bsnm{{MacKinnon}}, \binits{A.L.}},
\bauthor{\bsnm{{Shih}}, \binits{A.Y.}},
\bauthor{\bsnm{{Lin}}, \binits{R.P.}}:
\byear{2008}a,
\batitle{{Coronal {$\gamma$}-Ray Bremsstrahlung from Solar Flare-accelerated
  Electrons}}.
\bjtitle{Astrophys. J.}
\bvolume{678},
\bfpage{L63}\,--\,\blpage{L66}.
doi:\doiurl{10.1086/588381}.
\end{barticle}
\endbibitem

\bibitem[\protect\citeauthoryear{{Krucker} \textit{et~al.}}{2008b}]{kru08a}
\begin{barticle}
\bauthor{\bsnm{{Krucker}}, \binits{S.}},
\bauthor{\bsnm{{Battaglia}}, \binits{M.}},
\bauthor{\bsnm{{Cargill}}, \binits{P.J.}},
\bauthor{\bsnm{{Fletcher}}, \binits{L.}},
\bauthor{\bsnm{{Hudson}}, \binits{H.S.}},
\bauthor{\bsnm{{MacKinnon}}, \binits{A.L.}},
\bauthor{\bsnm{{Masuda}}, \binits{S.}},
\bauthor{\bsnm{{Sui}}, \binits{L.}},
\bauthor{\bsnm{{Tomczak}}, \binits{M.}},
\bauthor{\bsnm{{Veronig}}, \binits{A.L.}},
\bauthor{\bsnm{{Vlahos}}, \binits{L.}},
\bauthor{\bsnm{{White}}, \binits{S.M.}}:
\byear{2008}b,
\batitle{{Hard X-ray emission from the solar corona}}.
\bjtitle{Astron. Astrophys. Rev.}
\bvolume{16},
\bfpage{155}\,--\,\blpage{208}.
doi:\doiurl{10.1007/s00159-008-0014-9}.
\end{barticle}
\endbibitem

\bibitem[\protect\citeauthoryear{{Krucker} \textit{et~al.}}{2010}]{kru10}
\begin{barticle}
\bauthor{\bsnm{{Krucker}}, \binits{S.}},
\bauthor{\bsnm{{Hudson}}, \binits{H.S.}},
\bauthor{\bsnm{{Glesener}}, \binits{L.}},
\bauthor{\bsnm{{White}}, \binits{S.M.}},
\bauthor{\bsnm{{Masuda}}, \binits{S.}},
\bauthor{\bsnm{{Wuelser}}, \binits{J.-P.}},
\bauthor{\bsnm{{Lin}}, \binits{R.P.}}:
\byear{2010},
\batitle{{Measurements of the Coronal Acceleration Region of a Solar Flare}}.
\bjtitle{Astrophys. J.}
\bvolume{714},
\bfpage{1108}\,--\,\blpage{1119}.
doi:\doiurl{10.1088/0004-637X/714/2/1108}.
\end{barticle}
\endbibitem

\bibitem[\protect\citeauthoryear{{Kucera} \textit{et~al.}}{1993}]{kuc93}
\begin{barticle}
\bauthor{\bsnm{{Kucera}}, \binits{T.A.}},
\bauthor{\bsnm{{Dulk}}, \binits{G.A.}},
\bauthor{\bsnm{{Kiplinger}}, \binits{A.L.}},
\bauthor{\bsnm{{Winglee}}, \binits{R.M.}},
\bauthor{\bsnm{{Bastian}}, \binits{T.S.}},
\bauthor{\bsnm{{Graeter}}, \binits{M.}}:
\byear{1993},
\batitle{{Multiple wavelength observations of an off-limb eruptive solar
  flare}}.
\bjtitle{Astrophys. J.}
\bvolume{412},
\bfpage{853}\,--\,\blpage{864}.
doi:\doiurl{10.1086/172967}.
\end{barticle}
\endbibitem

\bibitem[\protect\citeauthoryear{{Kuznetsov} and {Zharkova}}{2010}]{kuz10}
\begin{barticle}
\bauthor{\bsnm{{Kuznetsov}}, \binits{A.A.}},
\bauthor{\bsnm{{Zharkova}}, \binits{V.V.}}:
\byear{2010},
\batitle{{Manifestations of Energetic Electrons with Anisotropic Distributions
  in Solar Flares. II. Gyrosynchrotron Microwave Emission}}.
\bjtitle{Astrophys. J.}
\bvolume{722},
\bfpage{1577}\,--\,\blpage{1588}.
doi:\doiurl{10.1088/0004-637X/722/2/1577}.
\end{barticle}
\endbibitem

\bibitem[\protect\citeauthoryear{{Kuznetsov}, {Nita}, and
  {Fleishman}}{2011}]{kuz11}
\begin{barticle}
\bauthor{\bsnm{{Kuznetsov}}, \binits{A.A.}},
\bauthor{\bsnm{{Nita}}, \binits{G.M.}},
\bauthor{\bsnm{{Fleishman}}, \binits{G.D.}}:
\byear{2011},
\batitle{{Three-dimensional Simulations of Gyrosynchrotron Emission from Mildly
  Anisotropic Nonuniform Electron Distributions in Symmetric Magnetic Loops}}.
\bjtitle{Astrophys. J.}
\bvolume{742},
\bfpage{87}.
doi:\doiurl{10.1088/0004-637X/742/2/87}.
\end{barticle}
\endbibitem

\bibitem[\protect\citeauthoryear{{Lee} and {Gary}}{2000}]{lee00}
\begin{barticle}
\bauthor{\bsnm{{Lee}}, \binits{J.}},
\bauthor{\bsnm{{Gary}}, \binits{D.E.}}:
\byear{2000},
\batitle{{Solar Microwave Bursts and Injection Pitch-Angle Distribution of
  Flare Electrons}}.
\bjtitle{Astrophys. J.}
\bvolume{543},
\bfpage{457}\,--\,\blpage{471}.
doi:\doiurl{10.1086/317080}.
\end{barticle}
\endbibitem

\bibitem[\protect\citeauthoryear{{Lin} \textit{et~al.}}{2002}]{lin02}
\begin{barticle}
\bauthor{\bsnm{{Lin}}, \binits{R.P.}},
\bauthor{\bsnm{{Dennis}}, \binits{B.R.}},
\bauthor{\bsnm{{Hurford}}, \binits{G.J.}},
\bauthor{\bsnm{{Smith}}, \binits{D.M.}},
\bauthor{\bsnm{{Zehnder}}, \binits{A.}},
\bauthor{\bsnm{{Harvey}}, \binits{P.R.}},
\bauthor{\bsnm{{Curtis}}, \binits{D.W.}},
\bauthor{\bsnm{{Pankow}}, \binits{D.}},
\bauthor{\bsnm{{Turin}}, \binits{P.}},
\bauthor{\bsnm{{Bester}}, \binits{M.}},
\bauthor{\bsnm{{Csillaghy}}, \binits{A.}},
\bauthor{\bsnm{{Lewis}}, \binits{M.}},
\bauthor{\bsnm{{Madden}}, \binits{N.}},
\bauthor{\bsnm{{van Beek}}, \binits{H.F.}},
\bauthor{\bsnm{{Appleby}}, \binits{M.}},
\bauthor{\bsnm{{Raudorf}}, \binits{T.}},
\bauthor{\bsnm{{McTiernan}}, \binits{J.}},
\bauthor{\bsnm{{Ramaty}}, \binits{R.}},
\bauthor{\bsnm{{Schmahl}}, \binits{E.}},
\bauthor{\bsnm{{Schwartz}}, \binits{R.}},
\bauthor{\bsnm{{Krucker}}, \binits{S.}},
\bauthor{\bsnm{{Abiad}}, \binits{R.}},
\bauthor{\bsnm{{Quinn}}, \binits{T.}},
\bauthor{\bsnm{{Berg}}, \binits{P.}},
\bauthor{\bsnm{{Hashii}}, \binits{M.}},
\bauthor{\bsnm{{Sterling}}, \binits{R.}},
\bauthor{\bsnm{{Jackson}}, \binits{R.}},
\bauthor{\bsnm{{Pratt}}, \binits{R.}},
\bauthor{\bsnm{{Campbell}}, \binits{R.D.}},
\bauthor{\bsnm{{Malone}}, \binits{D.}},
\bauthor{\bsnm{{Landis}}, \binits{D.}},
\bauthor{\bsnm{{Barrington-Leigh}}, \binits{C.P.}},
\bauthor{\bsnm{{Slassi-Sennou}}, \binits{S.}},
\bauthor{\bsnm{{Cork}}, \binits{C.}},
\bauthor{\bsnm{{Clark}}, \binits{D.}},
\bauthor{\bsnm{{Amato}}, \binits{D.}},
\bauthor{\bsnm{{Orwig}}, \binits{L.}},
\bauthor{\bsnm{{Boyle}}, \binits{R.}},
\bauthor{\bsnm{{Banks}}, \binits{I.S.}},
\bauthor{\bsnm{{Shirey}}, \binits{K.}},
\bauthor{\bsnm{{Tolbert}}, \binits{A.K.}},
\bauthor{\bsnm{{Zarro}}, \binits{D.}},
\bauthor{\bsnm{{Snow}}, \binits{F.}},
\bauthor{\bsnm{{Thomsen}}, \binits{K.}},
\bauthor{\bsnm{{Henneck}}, \binits{R.}},
\bauthor{\bsnm{{McHedlishvili}}, \binits{A.}},
\bauthor{\bsnm{{Ming}}, \binits{P.}},
\bauthor{\bsnm{{Fivian}}, \binits{M.}},
\bauthor{\bsnm{{Jordan}}, \binits{J.}},
\bauthor{\bsnm{{Wanner}}, \binits{R.}},
\bauthor{\bsnm{{Crubb}}, \binits{J.}},
\bauthor{\bsnm{{Preble}}, \binits{J.}},
\bauthor{\bsnm{{Matranga}}, \binits{M.}},
\bauthor{\bsnm{{Benz}}, \binits{A.}},
\bauthor{\bsnm{{Hudson}}, \binits{H.}},
\bauthor{\bsnm{{Canfield}}, \binits{R.C.}},
\bauthor{\bsnm{{Holman}}, \binits{G.D.}},
\bauthor{\bsnm{{Crannell}}, \binits{C.}},
\bauthor{\bsnm{{Kosugi}}, \binits{T.}},
\bauthor{\bsnm{{Emslie}}, \binits{A.G.}},
\bauthor{\bsnm{{Vilmer}}, \binits{N.}},
\bauthor{\bsnm{{Brown}}, \binits{J.C.}},
\bauthor{\bsnm{{Johns-Krull}}, \binits{C.}},
\bauthor{\bsnm{{Aschwanden}}, \binits{M.}},
\bauthor{\bsnm{{Metcalf}}, \binits{T.}},
\bauthor{\bsnm{{Conway}}, \binits{A.}}:
\byear{2002},
\batitle{{The Reuven Ramaty High-Energy Solar Spectroscopic Imager (RHESSI)}}.
\bjtitle{Solar Phys.}
\bvolume{210},
\bfpage{3}\,--\,\blpage{32}.
doi:\doiurl{10.1023/A:1022428818870}.
\end{barticle}
\endbibitem

\bibitem[\protect\citeauthoryear{{Melnikov}}{2006}]{mel06}
\begin{bchapter}
\bauthor{\bsnm{{Melnikov}}, \binits{V.F.}}:
\byear{2006},
\bctitle{{Electron Acceleration and Transport in Microwave Flaring Loops}}.
In: \beditor{\bsnm{{Bastian}}, \binits{T.}},
\beditor{\bsnm{{Gopalswamy}}, \binits{N.}},
\beditor{\bsnm{{Kosugi}}, \binits{T.}},
\beditor{\bsnm{{Shibasaki}}, \binits{K.}},
\beditor{\bsnm{{Yokoyama}}, \binits{T.}} (eds.)
\bbtitle{Solar Physics with the Nobeyama Radioheliograph},
\bfpage{11}\,--\,\blpage{22}.
\end{bchapter}
\endbibitem

\bibitem[\protect\citeauthoryear{{Nakagawa} and {Raadu}}{1972}]{nak72}
\begin{barticle}
\bauthor{\bsnm{{Nakagawa}}, \binits{Y.}},
\bauthor{\bsnm{{Raadu}}, \binits{M.A.}}:
\byear{1972},
\batitle{{On Practical Representation of Magnetic Field}}.
\bjtitle{Solar Phys.}
\bvolume{25},
\bfpage{127}\,--\,\blpage{135}.
doi:\doiurl{10.1007/BF00155751}.
\end{barticle}
\endbibitem

\bibitem[\protect\citeauthoryear{{Nakajima} \textit{et~al.}}{1994}]{nak94}
\begin{barticle}
\bauthor{\bsnm{{Nakajima}}, \binits{H.}},
\bauthor{\bsnm{{Nishio}}, \binits{M.}},
\bauthor{\bsnm{{Enome}}, \binits{S.}},
\bauthor{\bsnm{{Shibasaki}}, \binits{K.}},
\bauthor{\bsnm{{Takano}}, \binits{T.}},
\bauthor{\bsnm{{Hanaoka}}, \binits{Y.}},
\bauthor{\bsnm{{Torii}}, \binits{C.}},
\bauthor{\bsnm{{Sekiguchi}}, \binits{H.}},
\bauthor{\bsnm{{Bushimata}}, \binits{T.}},
\bauthor{\bsnm{{Kawashima}}, \binits{S.}},
\bauthor{\bsnm{{Shinohara}}, \binits{N.}},
\bauthor{\bsnm{{Irimajiri}}, \binits{Y.}},
\bauthor{\bsnm{{Koshiishi}}, \binits{H.}},
\bauthor{\bsnm{{Kosugi}}, \binits{T.}},
\bauthor{\bsnm{{Shiomi}}, \binits{Y.}},
\bauthor{\bsnm{{Sawa}}, \binits{M.}},
\bauthor{\bsnm{{Kai}}, \binits{K.}}:
\byear{1994},
\batitle{{The Nobeyama radioheliograph.}}
\bjtitle{IEEE Proc.}
\bvolume{82},
\bfpage{705}\,--\,\blpage{713}.
\end{barticle}
\endbibitem

\bibitem[\protect\citeauthoryear{{Nindos} \textit{et~al.}}{2000}]{nin00}
\begin{barticle}
\bauthor{\bsnm{{Nindos}}, \binits{A.}},
\bauthor{\bsnm{{White}}, \binits{S.M.}},
\bauthor{\bsnm{{Kundu}}, \binits{M.R.}},
\bauthor{\bsnm{{Gary}}, \binits{D.E.}}:
\byear{2000},
\batitle{{Observations and Models of a Flaring Loop}}.
\bjtitle{Astrophys. J.}
\bvolume{533},
\bfpage{1053}\,--\,\blpage{1062}.
doi:\doiurl{10.1086/308705}.
\end{barticle}
\endbibitem

\bibitem[\protect\citeauthoryear{{Nita} \textit{et~al.}}{2011a}]{nit11a}
\begin{bchapter}
\bauthor{\bsnm{{Nita}}, \binits{G.M.}},
\bauthor{\bsnm{{Fleishman}}, \binits{G.D.}},
\bauthor{\bsnm{{Gary}}, \binits{D.E.}},
\bauthor{\bsnm{{Kuznetsov}}, \binits{A.A.}},
\bauthor{\bsnm{{Kontar}}, \binits{E.P.}}:
\byear{2011}a,
\bctitle{{GX\_Simulator: An Interactive Idl Widget Tool For Visualization And
  Simulation Of Imaging Spectroscopy Models And Data}}.
In: \bbtitle{Bull. Am. Astron. Soc.}
\bseriesno{43},
\bfpage{1811}.
\end{bchapter}
\endbibitem

\bibitem[\protect\citeauthoryear{{Nita} \textit{et~al.}}{2011b}]{nit11b}
\begin{botherref}
\oauthor{\bsnm{{Nita}}, \binits{G.M.}},
\oauthor{\bsnm{{Fleishman}}, \binits{G.D.}},
\oauthor{\bsnm{{Gary}}, \binits{D.E.}},
\oauthor{\bsnm{{Kuznetsov}}, \binits{A.}},
\oauthor{\bsnm{{Kontar}}, \binits{E.P.}}:
2011b,
{Novel 3D Approach to Flare Modeling via Interactive IDL Widget Tools}.
\textit{AGU Fall Meeting 2011},
A7.
\end{botherref}
\endbibitem

\bibitem[\protect\citeauthoryear{{Nita} \textit{et~al.}}{2012}]{nit12}
\begin{bchapter}
\bauthor{\bsnm{{Nita}}, \binits{G.M.}},
\bauthor{\bsnm{{Fleishman}}, \binits{G.D.}},
\bauthor{\bsnm{{Gary}}, \binits{D.E.}},
\bauthor{\bsnm{{Kuznetsov}}, \binits{A.A.}},
\bauthor{\bsnm{{Kontar}}, \binits{E.P.}}:
\byear{2012},
\bctitle{{Integrated Idl Tool For 3d Modeling And Imaging Data Analysis}}.
In: \bbtitle{AAS Meeting}
\bseriesno{220},
\bfpage{204{.}51}.
\end{bchapter}
\endbibitem

\bibitem[\protect\citeauthoryear{{Preka-Papadema} and
  {Alissandrakis}}{1992}]{pre92}
\begin{barticle}
\bauthor{\bsnm{{Preka-Papadema}}, \binits{P.}},
\bauthor{\bsnm{{Alissandrakis}}, \binits{C.E.}}:
\byear{1992},
\batitle{{Two-dimensional model maps of flaring loops at cm-wavelengths}}.
\bjtitle{Astron. Astrophys.}
\bvolume{257},
\bfpage{307}\,--\,\blpage{314}.
\end{barticle}
\endbibitem

\bibitem[\protect\citeauthoryear{{Seehafer}}{1978}]{see78}
\begin{barticle}
\bauthor{\bsnm{{Seehafer}}, \binits{N.}}:
\byear{1978},
\batitle{{Determination of constant alpha force-free solar magnetic fields from
  magnetograph data}}.
\bjtitle{Solar Phys.}
\bvolume{58},
\bfpage{215}\,--\,\blpage{223}.
doi:\doiurl{10.1007/BF00157267}.
\end{barticle}
\endbibitem

\bibitem[\protect\citeauthoryear{{Tzatzakis}, {Nindos}, and
  {Alissandrakis}}{2008}]{tza08}
\begin{barticle}
\bauthor{\bsnm{{Tzatzakis}}, \binits{V.}},
\bauthor{\bsnm{{Nindos}}, \binits{A.}},
\bauthor{\bsnm{{Alissandrakis}}, \binits{C.E.}}:
\byear{2008},
\batitle{{A Statistical Study of Microwave Flare Morphologies}}.
\bjtitle{Solar Phys.}
\bvolume{253},
\bfpage{79}\,--\,\blpage{94}.
doi:\doiurl{10.1007/s11207-008-9263-z}.
\end{barticle}
\endbibitem

\bibitem[\protect\citeauthoryear{{Wang} \textit{et~al.}}{1995}]{wan95}
\begin{barticle}
\bauthor{\bsnm{{Wang}}, \binits{H.}},
\bauthor{\bsnm{{Gary}}, \binits{D.E.}},
\bauthor{\bsnm{{Zirin}}, \binits{H.}},
\bauthor{\bsnm{{Schwartz}}, \binits{R.A.}},
\bauthor{\bsnm{{Sakao}}, \binits{T.}},
\bauthor{\bsnm{{Kosugi}}, \binits{T.}},
\bauthor{\bsnm{{Shibata}}, \binits{K.}}:
\byear{1995},
\batitle{{Coordinated OVRO, BATSE, Yohkoh, and BBSO Observations of the 1992
  June 25 M1.4 Flare}}.
\bjtitle{Astrophys. J.}
\bvolume{453},
\bfpage{505}.
doi:\doiurl{10.1086/176411}.
\end{barticle}
\endbibitem

\end{thebibliography}
\end{document}